\documentclass[11pt,a4paper]{article}
\pdfoutput=1
\usepackage[margin=1.25in]{geometry}
\usepackage{amsmath}
\usepackage{graphicx}
\graphicspath{{./}{./Figures/}}
\PassOptionsToPackage{hyphens}{url}\usepackage[hidelinks]{hyperref}
\usepackage{float} 
\usepackage{latexsym}
\usepackage{libertine}
\usepackage[libertine]{newtxmath}
\usepackage[font=footnotesize,labelfont=bf]{caption}
\usepackage[T1]{fontenc}
\usepackage{textgreek}
\usepackage[utf8]{inputenc}
\usepackage{mathrsfs}
\usepackage{booktabs}
\usepackage[numbers]{natbib}
\usepackage{a4wide}
\usepackage{makecell}
\usepackage{xcolor}

\title{Building population models for large-scale neural recordings: opportunities and pitfalls}
\author{
Cole Hurwitz*, Nina Kudryashova*, Arno Onken and Matthias H. Hennig$^\dagger$\\
\small {University of Edinburgh, Institute for Adaptive and Neural Computation}\\
\small{Edinburgh, EH8 9AB, United Kingdom}\\
\footnotesize{$\dagger$ Correspondence: \texttt{m.hennig@ed.ac.uk}}\\
\footnotesize{* equal contribution}
}
\date{April, 2021}

\begin{document}

\maketitle

\abstract{Modern recording technologies now enable simultaneous recording from large numbers of neurons. This has driven the development of new statistical models for analyzing and interpreting neural population activity. Here we provide a broad overview of recent developments in this area. We compare and contrast different approaches, highlight strengths and limitations, and discuss biological and mechanistic insights that these methods provide.} 

\section*{Introduction}


Large scale recordings from neurons with dense, high channel count probes now yield sufficient data to understand the precise covariation in neural populations \cite{junFullyIntegratedSilicon2017, raducanu2017time,saxena2019towards}. This departure from analyzing the tuning properties of individual neuron, often pooled across experiments, requires model-based approaches that capture the complexity of high-dimensional population activity while remaining computationally tractable.


Statistical models have emerged as an essential tool for overcoming this challenge. 
Informally, these models can be grouped into two categories: fully observed models and latent variable models. Fully observed models try to explicitly capture the interactions between neurons by directly modeling the joint activity of the population. Since modeling the full space of activity patterns is intractable, a key ingredient for tractable fully observed models is an efficient description of the interactions between neurons that retains essential characteristics.
Latent variable models, on the other hand, assume that population activity can be summarized by a small set of variables called latent factors. Rather than directly modeling interactions, latent variable models capture neuronal interactions through activation of these factors.

This conceptual difference means that the specific hypotheses each of these model classes can address are also different and that conclusions drawn from such analysis have to be considered in the light of the main assumptions the model is based upon. In this review, we first briefly discuss how single neuron activity is extracted from high-density recordings. Next, we provide an overview of current developments and show which questions can and cannot be addressed with each model class. We close with a discussion of limitations and future directions.

\section*{Extracting modeling data}

\begin{figure}[t]
    \centering
    \includegraphics[width=0.9\textwidth]{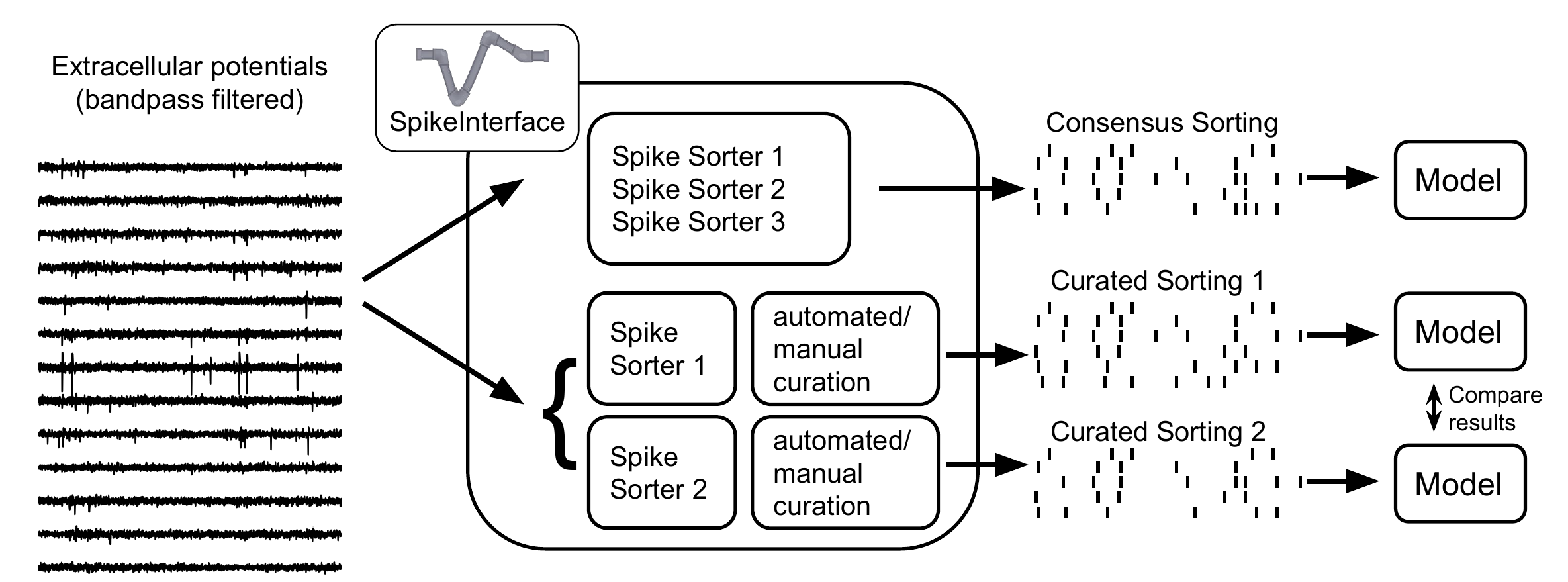}
    \caption{The process of extracting the activity of single neurons with a spike sorter is error-prone with different sorters having different failure modes. It is, therefore, recommended to make use of different sorters before performing model-based analysis. A first method is to perform the same analysis on the output of different sorters to see if functional insights are independent of the chosen spike sorter (bottom). Alternatively, one could obtain a consensus sorting (top) where false positive units are largely removed by taking the agreement between different sorters. Any analyses could then be performed on this consensus sorting. These pipelines, together with various pre- and postprocessing steps as well as automated curation, can be easily implemented using the SpikeInterface framework \cite{buccinoSpikeInterfaceUnifiedFramework2020} which supports all commonly used spike sorters.}
    \label{fig:spikesort}
\end{figure}

Large scale recordings can be obtained with high density extracellular probes or using calcium imaging. From extracellular recordings, spike trains of individual neurons are extracted through a process called \textit{spike sorting}. 
There are several algorithmic approaches for spike sorting of data from dense probes which are reviewed elsewhere \cite{reyPresentFutureSpike2015,hennigScalingSpikeDetection2019,carlsonContinuingProgressSpike2019,lefebvreRecentProgressMultielectrode2016}. Accurately extracting spike trains is essential since mistakes, including incorrectly merged or split units, unresolved overlapping spikes, or incorrect clustering due to probe drift, have been shown to bias subsequent analysis \cite{bar2001failure, ventura2009traditional, ventura2012accurately}. 
For example, in regions with high-firing rates and synchronicity, spike sorting mistakes can lead to artificial correlations between independently firing neurons \cite{bar2001failure, ecker2010decorrelated}. 

It is important to acknowledge that none of the existing spike sorting algorithms are error-free, and that the scale of these recordings now makes manual curation difficult. To aid selection of an appropriate spike sorter, a large collection of ground truth comparisons is available on the SpikeForest website \cite{maglandSpikeForestReproducibleWebfacing2020}. Moreover, the comparison of spike sorter outputs has shown surprisingly low agreement in part due to a large number of false positive units \cite{buccinoSpikeInterfaceUnifiedFramework2020}. A viable approach to mitigate against spurious findings is, therefore, to perform the same analysis on multiple sorter outputs or on an ensemble agreement among multiple sorters; this can be done using a software framework such as SpikeInterface (Figure~\ref{fig:spikesort}) \cite{buccinoSpikeInterfaceUnifiedFramework2020}.

While calcium imaging allows unequivocal identification of single neurons, the much lower sampling rate and the binding kinetics of calcium indicators mean the recorded signal may be an incomplete record of the spiking activity of these neurons \cite{huang2021relationship}. A first approach is to recover neural spike trains using deconvolution methods. Similar to spike sorting however, different algorithms have been shown to produce conflicting results when analyzing correlations between neurons \cite{evans2019use}. It is therefore prudent to avoid analysis based only on a single convolution algorithm. Second, recent work aims to to include a model of the dynamics giving rise to the calcium signal directly into statistical models \cite{prince2021parallel}. This approach is promising in particular for latent variable models if the different sources of variation (neural and calcium dynamics) can be effectively disentangled.

\section*{Fully observed models}
\begin{figure}[t]
    \centering
    \includegraphics[width=0.7\textwidth]{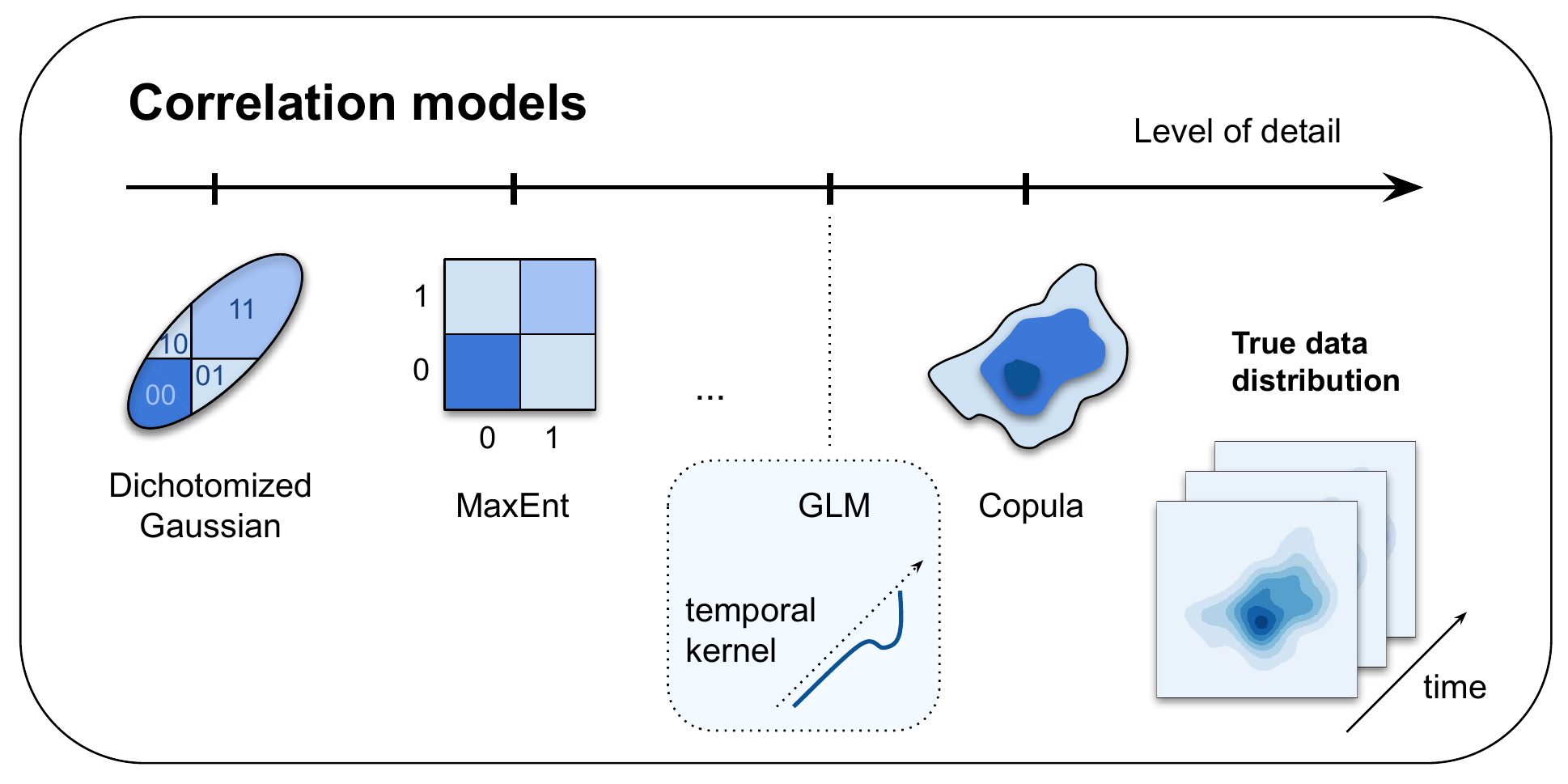}
    \caption{Fully observed models construct a joint distribution of population activity with various levels of detail. The horizontal arrow orders the methods according to their expressivity (i.e. how much dependence structure the method can account for).}
    \label{fig:corr}
\end{figure}

When recording from just 20 neurons, there are over one million possible instantaneous ON/OFF patterns of spiking ($2^{20}$) for a small time bin; this number grows to one billion for 30 neurons ($2^{30}$). While the number of spike patterns that are actually observed is much smaller and constrained by the connectivity between neurons, estimating all pattern frequencies reliably from typical recordings is impossible even for small populations. Fully observed statistical models are a tool to achieve precisely this by approximating this full distribution using as few parameters as possible together with suitable constraints \cite{averbeck2006neural,frankeStructuresNeuralCorrelation2016,zylberbergDirectionSelectiveCircuits2016}. As always, different approaches exist with different trade-offs between complexity, tractability and interpretability~(see~Fig.~\ref{fig:corr} and Tab.~\ref{table:fullobs}). A main purpose of these models is to ask how much information recorded activity contains about stimuli or behavior, which can be addressed using information-theoretical or decoding approaches.


\subsubsection*{Models}

\textit{Maximum Entropy models} (MaxEnt) provide a principled way to construct probability distributions for spiking activity by finding the least structured distribution that fulfills specific constraints obtained from the data~\cite{savin2017maximum}. Typical constraints for MaxEnt models are the firing rates of and pairwise correlations between the neurons~\cite{schneidmanWeakPairwiseCorrelations2006,shlensStructureMultineuronFiring2006, tkacikSimplestMaximumEntropy2013} and the population synchrony ~\cite{tkacikSimplestMaximumEntropy2013}.
While MaxEnt models have been shown to model spiking activity well, the computational cost of fitting them is prohibitive for large neuronal populations. This has led to the development of analytically tractable models constrained by the dependencies between individual neurons and the aggregate population activity~\cite{tkacikSimplestMaximumEntropy2013,gardellaTractableMethodDescribing2016,okun2015diverse,odonnellPopulationTrackingModel2017}.
Related and more scalable is the \textit{Dichotomized Gaussian model}
~\cite{amariStochasticMechanismSynchronous2003,mackeGeneratingSpikeTrains2009} which can be used to model spike count variables with arbitrary marginal distributions~\cite{mackeGeneratingSpikeTrains2009}.

MaxEnt models cannot easily take external inputs into account. If these are well specified, \textit{generalized linear models} (GLMs) can successfully describe the activity of large populations~\cite{truccoloPointProcessFramework2005,pillowPredictionDecodingRetinal2005,pillowSpatiotemporalCorrelationsVisual2008}.
In these models, external inputs, inputs from other neurons, and each neuron's spiking history are weighted, summed up and transformed by a rate function, which drives a stochastic process to model spiking activity.
For typical rate functions, fitting GLMs is a convex optimization problem which can be solved efficiently, making GLMs scalable to large-scale neural recordings~\cite{truccoloPointProcessFramework2005,pillowSpatiotemporalCorrelationsVisual2008}.
Along with modeling the observed neural population, the activity of unobserved neurons can be inferred \cite{battistin2015belief}, and knowing the topology of the external covariates can further aid in inferring the activity of the unseen neurons~\cite{spreemann2018using}.
Despite their flexibility, GLMs cannot model networks with neurons that perform non-linear integration of multiple external inputs~\cite{benjamin2018modern}.

A more direct representation of statistical dependencies is provided by \textit{copula models}, which decompose neural population activity into the distributions of individual neuron activities and the dependence between these neurons, represented by a copula~\cite{jenisonShapeNeuralDependence2004}.
Copula models can represent inter-neuronal dependencies that go beyond linear (Gaussian) correlations \cite{berkes2008characterizing}, as well as non-linear stimulus-response relationships~\cite{kudryashova2020parametric}.
These highly flexible models can for example be used to jointly model different recording modalities such as local field potentials and individual neurons' spiking activities~\cite{aas2009pair,onkenMixedVineCopulas2016}.
They are particularly promising for analyzing recordings of neural data and exploring its relation to behavior and stimuli which typically follow different statistics and tend to have complex dependencies~\cite{kudryashova2020parametric,ince2017statistical}.

\subsubsection*{Interpretation and biological insights}

\begin{table}[t]
\centering
\begin{tabular}{m{2.5cm} m{3.5cm} | c c c c} 
\hline\hline 
Model & References & \shortstack{Number of\\ parameters} & \shortstack{Closed-form \\ pattern probabilities?} & \shortstack{Fit for\\ large $N$?} &  \\ [0.5ex] 
\hline 
Dichotomized Gaussian & \shortstack{\citet{amariStochasticMechanismSynchronous2003} \\ \citet{mackeGeneratingSpikeTrains2009}} & $\sim N^2$ & No & Yes \\ [10pt] 
\shortstack{Pairwise \\ MaxEnt} & \shortstack{\citet{schneidmanWeakPairwiseCorrelations2006} \\ \citet{shlensStructureMultineuronFiring2006}} & $\sim N^2$ & No$^1$ & Difficult \\ [12pt]  
\shortstack{Tractable\\ MaxEnt} & \shortstack{\citet{tkacikSimplestMaximumEntropy2013} \\ \citet{gardellaTractableMethodDescribing2016} \\ \citet{odonnellPopulationTrackingModel2017}} & $\sim N^2$ & Yes$^1$ & Yes \\ [16pt]  
GLM & \citet{pillowSpatiotemporalCorrelationsVisual2008} & $\sim DN^2$ & No & Difficult \\ [6pt] 
\shortstack{Vine \\ Copula} & \shortstack{\citet{aas2009pair}\\ \citet{onkenMixedVineCopulas2016}}  & $\sim DN^2 $ & Yes & Yes \\ [6pt] 
\hline 
\end{tabular}
\caption{Fully observed models: A table (adapted from~\citet{odonnellPopulationTrackingModel2017}) characterizing model properties and limitations. Here, $N$ is the number of neurons and $D$ is the number of coefficients per interaction term, such as filter sizes for GLM or number of parameters for parametric copula families. Sampling is possible from all of these models. 1. For a more detailed comparison of MaxEnt models, see Table~1 in~\cite{savin2017maximum}.
} 
\label{table:fullobs} 
\end{table}

Fully observed models can be used in two main areas. First, they provide access to the full joint activity distribution of the population in place of direct, typically biased estimates from the recordings for information theoretic analyses and for decoding~\cite{ince2017statistical, kudryashova2020parametric,runyanDistinctTimescalesPopulation2017,pillow2011model}.
GLMs and copula models that were fit to population activity have been shown to have high decoding performance of external variables ~\cite{truccoloPointProcessFramework2005,pillowSpatiotemporalCorrelationsVisual2008,kulkarni2007common,lawhern2010population,ince2017statistical,kudryashova2020parametric}. For instance, GLMs with coupling filters were shown to capture 40\% more visual information from the retina than optimal linear decoding~\cite{pillowSpatiotemporalCorrelationsVisual2008}, indicating that GLMs can model additional details in the activity that are relevant for representing the stimulus.
In large-scale datasets with complex statistics, GLMs may however not be appropriate. In these cases, copula models can be useful for measuring the amount of information in the population response about external covariates~\cite{onkenMixedVineCopulas2016, ince2017statistical, kudryashova2020parametric}.
Additionally, for complex stimuli with spatial or temporal correlations, one must distinguish `stimulus correlations' that arise from stimulus statistics from `noise correlations' due to circuit interactions. Including this yields better performance in MaxEnt models~\cite{granot2013stimulus}, and prevents spurious effects in GLMs~\cite{mahuas2020new}.
More generally, comparisons of these models with linear decoders can quantify the information that emerges from the network interactions, i.e. how the whole network is different from the sum of its parts~\cite{pillowPredictionDecodingRetinal2005,pillowSpatiotemporalCorrelationsVisual2008,schneidmanWeakPairwiseCorrelations2006,ohiorhenuanSparseCodingHighorder2010}. 

Second, the parameters in many fully observed models can be interpreted as the strength of the interactions between neurons. For example, the coupling parameters in dichotomized Gaussian and in pairwise MaxEnt models describe the strength of pairwise interactions between single neurons~\cite{mackeGeneratingSpikeTrains2009,schneidmanWeakPairwiseCorrelations2006}. As these connections are inferred from the population activity alone, they are usually referred to as functional interactions or connections, to emphasize the difference to synaptic connections. This analysis has been used to uncover the spatial extent of interactions in the retina~\cite{shlensStructureMultineuronFiring2006, ohiorhenuanSparseCodingHighorder2010} and to characterize interactions between neurons with different selectivity in the entorhinal cortex~\cite{dunnCorrelationsFunctionalConnections2015}. 

Whether or not inferred functional connectivity is biologically interpretable critically depends on the identifiability of the relevant parameters, and here caution should be exercised. For instance, while the coupling filters in GLMs have unique solutions, filter parameter changes do not necessarily change the GLM responses much~\cite{brinkman2018predicting} suggesting that, in practice, coupling filters may not be properly identifiable. Similarly, in MaxEnt models inferred functional connectivity is not well constrained by the recorded spike trains. As a result, many coupling parameters, even those with large absolute values, can be altered without significantly changing the network activity predicted by the model~\cite{panas2015sloppiness,ponce-alvarezCorticalStateTransitions2020}. 

Limited data can be a main contributor to poor parameter identifiability. To address this, the uncertainty of the model parameters
can be quantified with approximate Bayesian inference methods~\cite{zoltowski2018scaling}. A recently developed LR-GLM utilized low-rank data approximations to scale approximate Bayesian inference methods to high-dimensional real datasets making these methods more applicable to large-scale neural recordings~\cite{trippe2019lr}. However, closer inspection of MaxEnt models suggests weakly constrained couplings could also be an intrinsic property of neural circuits. Simulations have shown that re-wiring even a large number of synapses may only have minor functional consequences~\cite{mongilloInhibitoryConnectivityDefines2018}, hence the corresponding inferred connections are also weakly specified.
Analysis of long term recordings has shown that these less important connections are also subject to more intensive re-modeling over time~\cite{panas2015sloppiness,ponce-alvarezCorticalStateTransitions2020}.
Therefore, taking into account the reliability of the inferred couplings in functional connectivity models allows for the identification of the neurons, connections and circuit motifs that most determine the population activity statistics~\cite{herzogDimensionalityReductionSpatioTemporal2018,panas2015sloppiness,ponce-alvarezCorticalStateTransitions2020}.




\begin{figure}[t]
    \centering
    \includegraphics[width=0.7\textwidth]{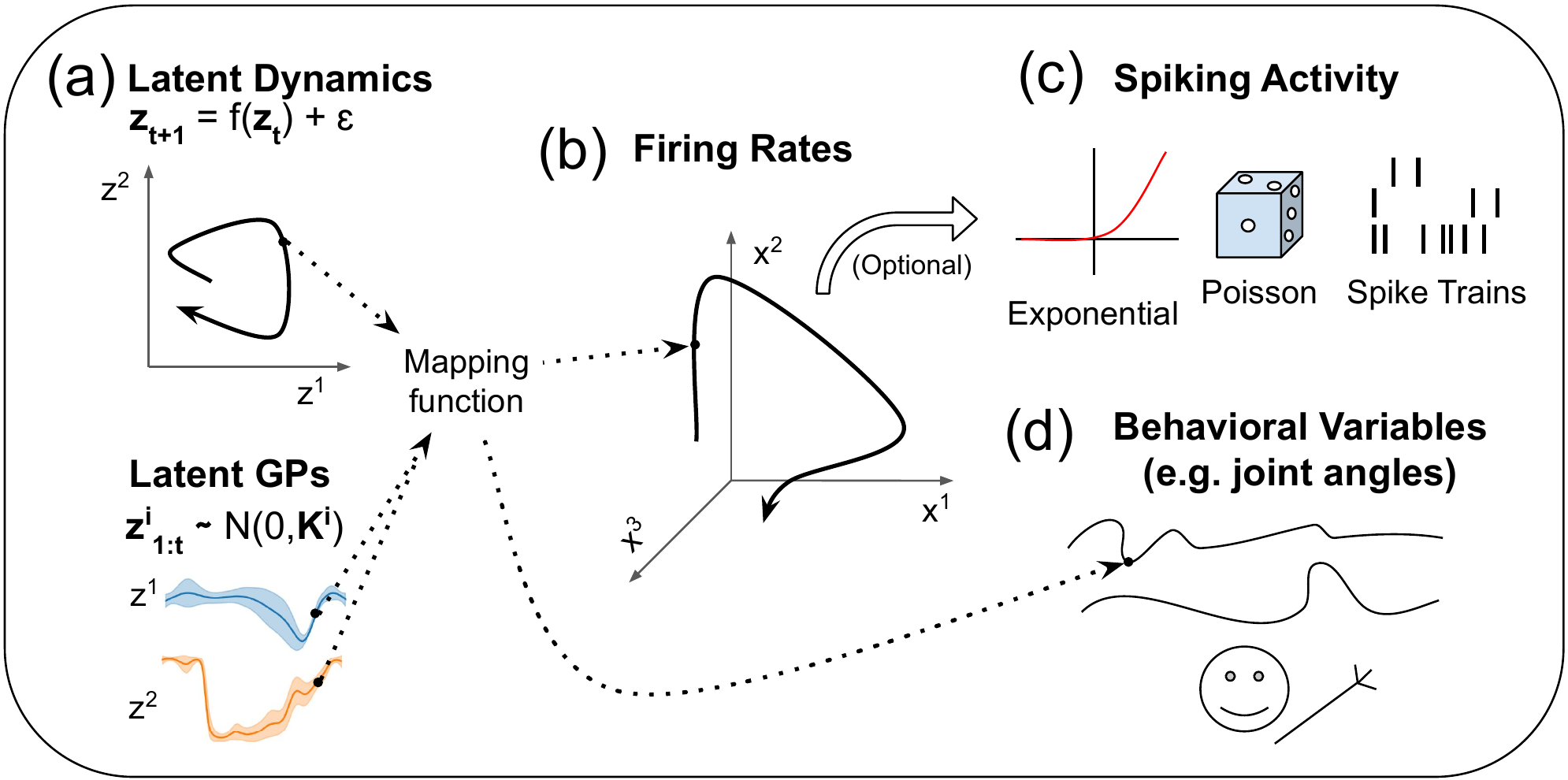}
    \caption{Latent variable models assume that neural activity is generated by a set of low-dimensional latent states. (a) When modeling the temporal evolution of the latent states, one can explicitly model the dynamics with a state-space model or can instead model the statistics of the latent trajectories with a Gaussian Process (GP). (b) The latent trajectories can be related back to neural activity (e.g. the firing rates) with a variety of mapping functions including linear functions, neural networks, and GPs. (c) An additional step to further constrain the dynamics is to reconstruct the spiking activity using a point-process distribution such as the Poisson distribution. (d) Recently, it has been shown that explicitly modeling the relationship between the latent factors and the observed behavioral variables can further constrain the factors to be low-dimensional and 'behaviorally relevant' \cite{saniModelingBehaviorallyRelevant2020}}
    \label{fig:latent}
\end{figure}

\section*{Latent variable models}

Despite the apparent high-dimensionality of population activity, recent studies suggest that the activity patterns underlying neural function are confined to low-dimensional manifolds \cite{mante2013context, churchland2012neural, sadtler2014neural, hall2014real, gallego2017neural, elsayed2017structure}. In other words, neural computation results from the activation of specific population-wide activity patterns rather than the independent modulation of single neurons \cite{gallegoLongtermStabilityCortical2020}. In this regime, latent variable models are a natural choice as they aim to capture arbitrarily complicated response structure in the neural activity with only a few latent variables. 


\subsection*{Models}

Latent variable models aim to capture the population activity structure through the temporal evolution of their learned latent variables. There are two main approaches for modeling the latent trajectories (shown in Figure~\ref{fig:latent} and summarized in Table~\ref{table:lvm}): \textit{State-space representations}, where the dynamics of the latent trajectories are modeled explicitly with an evolution function that relates past latent states to future latent states, and \textit{moment representations} where, rather than having an explicit model of the dynamics, the statistics of the latent trajectories over time are modeled. With both approaches, there is an \textit{observation model} which relates the neural trajectories back to the observed activity.

Traditional dimensionality reduction techniques such as principal component analysis (PCA) and factor analysis (FA) can be re-interpreted as \textit{static} state-space models without temporal dependence between states \cite{roweis1999unifying}. 
Extending these, \textit{linear} dynamical state-space models such as
linear dynamical systems (LDS) \cite{smith2003estimating, kulkarni2007common, buesing2013spectral, pfau2013robust} and jPCA
\cite{churchland2012neural, shenoy2013cortical, nemati2014probabilistic} can capture simple temporal dependencies between latent states. 
While efficient, these methods cannot model non-linear dynamics which are thought to underlie rhythmic motor patterns \cite{russo2020neural,hall2014common}, decision making \cite{rabinovich2008transient}, and pathologies such as epilepsy \cite{haghighi2017new}. To capture these dynamics, \textit{non-linear} dynamical state-space models have been introduced including approaches that model dynamics with recurrent neural networks (RNNs) \cite{pandarinathInferringSingletrialNeural2018, she2020neural}, piecewise-linear RNNs \cite{durstewitz2017state}, switching linear dynamics \cite{petreskaDynamicalSegmentationSingle2011}, and recurrent switching linear dynamics \cite{linderman2017bayesian, zoltowski2020general}. A particularly efficient RNN-based model is the recently introduced latent factor analysis via dynamical systems (LFADS) which utilizes neural networks to reduce the computational cost of inferring dynamics, making it more practical for large-scale neural recordings \cite{pandarinathInferringSingletrialNeural2018}. 
Despite these recent developments, non-linear state-space models are typically computationally expensive and parameter inference is difficult \cite{rutten2020non}.

An alternative to explicitly learning a dynamical model for the neural population is to model a statistical representation of the trajectories, typically as a Gaussian process (GP). GPs define a distribution over functions and allow to constrain certain features of the function such as its smoothness. Unlike state-space models, GP-based models provide both uncertainty quantification and principled model selection. The most common GP-based latent variable model is the Gaussian-process factor analysis (GPFA) \citep{byron2009gaussian}. In this model, each trajectory of the latent state is sampled from an independent GP.
This independence constraint, however, limits the ability of GPFA to model joint temporal dependencies making them less useful for uncovering temporal structure in neural data; for instance, GPFA cannot disentangle motor cortex trajectories during hand reach~\cite{rutten2020non}. 
Gaussian Process Factor Analysis with Dynamical Structure (GPFADS) seeks to remedy this by constraining the learned trajectories to have lower probabilities of occurring in reverse~\cite{rutten2020non}.
As a result, GPFADS can disentangle the trajectories generated by non-linear non-reversible dynamical systems such as primary motor cortex arm reach data. Overall, while GP-based latent variable models provide useful analytical properties and are generally quite efficient, they still need to demonstrate that they can capture the complex temporal dynamics that underlie neural activity as well as state-space models can.


An important aspect of all latent variable models is the observation model which relates the trajectories back to the neural activity. Common approaches include modeling neural activity with a conditional Gaussian model or the more plausible conditional Poisson model~\cite{macke2012empirical}.
The mapping between the latent trajectories and neural activity distribution can take on a number of functional forms including linear maps \cite{smith2003estimating, byron2009gaussian, pandarinathInferringSingletrialNeural2018, zhao2017variational}, GPs \cite{lawrence2005probabilistic, wu2017gaussian, she2020neural}, or neural networks \cite{gao2016linear}. Most models, however, rely on a simple linear mapping function as it forces the latent variables to capture more information about the joint neural activity.
Recently, observation models that take into account external behavioral variables have been introduced for capturing `behaviorally relevant' latent trajectories. \citet{saniModelingBehaviorallyRelevant2020} introduced a novel linear Gaussian state-space model (PSID) that jointly models the neural activity and the recorded behavioral variables. 
Intriguingly, PSID found that behaviorally relevant latent trajectories are significantly lower-dimensional than previously thought.

\subsubsection*{Interpretation and biological insights}


Unlike fully observed models, latent variable models provide a very succinct description of joint population activity in the form of latent trajectories. 
A simple yet effective approach to interpreting these latent trajectories is visualization; reducing the number of latent variables down to just two or three allows for insightful visualizations. For example, \citet{santhanam2009factor} used factor analysis to reduce preparatory neural activity in the premotor cortex to just three dimensions that, when visualized, provided evidence that the population activity contained information that could discriminate between goal-related target conditions in a delayed center-out reach task. 

It may however not be appropriate to reduce the number of latent variables to two or three for visualization without sacrificing model fit\footnote{Model fit can be measured in different ways including the amount of variance explained or the cross-validated likelihood of observed activity given the model. GP-based latent variable models can also provide uncertainty estimates for the latent trajectories.}.
A more principled and flexible way to interpret the learned latent trajectories is to relate them back to a measured external variable \cite{whiteway2019quest}. 
For example, one can measure how informative a latent variable is about an external variable by decoding the external variable at each time step using a simple method such as a least squares regression. 
While this approach has been used to interpret population response structure in the olfactory and visual system \cite{saha2013spatiotemporal, ecker2014state, si2019structured, zhao2016interpretable, stringerHighdimensionalGeometryPopulation2019}, its value is best demonstrated in studies that correlate latent trajectories in the premotor and motor cortex to arm reaching tasks.
\citet{churchland2012neural} found that low-dimensional rotational trajectories in the primary motor cortex were correlated with arm reaches. \citet{gallegoLongtermStabilityCortical2020} extended this finding, demonstrating that these low-dimensional trajectories are a stable neural correlate for consistent execution of arm reaches over the course of many years. Moreover, \citet{ramanathan2018low} showed that following a stroke, diminished reaching function in the contralesional arm was correlated with a loss of motor cortical neural trajectories. The trajectories only reemerged after motor recovery and proved to be a useful neuromodulatory target for therapeutic electrical stimulation. 

Despite the success of latent variable models, their usefulness can be dependent on the properties of the neuronal population and the specific experimental setting.
Firstly, latent variable models are most useful when the neural population activity is, in fact, low-dimensional. While a number of studies have independently found this to be the case for their datasets, recent work by \citet{stringerHighdimensionalGeometryPopulation2019} suggests that the manifold dimensionality of stimulus-evoked activity in the visual cortex is actually as high as it can be without becoming non-differentiable. Moreover, \citet{gaoTheoryMultineuronalDimensionality2017} suggests that population activity is as high-dimensional as possible given the simplicity of the given stimuli or tasks. While latent variable models can still be used to model this data (by increasing the number of latent variables), they may no longer be as interpretable. 
Another assumption in some models is that the latent trajectories are autonomous\footnote{\citet{pandarinathInferringSingletrialNeural2018} showed that LFADS could capture task-related inputs to a neural population by incorporating them into the generative process. It is still an open question how generalizable this approach is to other types of inputs.}, i.e. they only depend on an initial condition (usually trial-specific) and an evolution function. This limits the ability of these models to capture unmeasured inputs to the neural population from other brain regions. Finally, and most importantly, latent variable models, by design, will mix all sources of neural variability in the latent space \cite{whiteway2019quest}. This can make interpreting the latent variables challenging in particular for complex behaviors or diverse stimulus ensembles. These constraints may be the reason why many of the conceptual breakthroughs provided by latent variable models are for motor cortical activity where the dimensionality appears low, the trajectories are largely autonomous given the trial conditions, and the variation is largely explained by the measured behavior. A possible way forward to improve the interpretability of these models in more complex experimental regimes is to explicitly model external variables either in the observation model \cite{kobak2016demixed, saniModelingBehaviorallyRelevant2020} or in the latent space \cite{zhou2020learning}.


\begin{table}[t]
\centering 
\begin{tabular}{c c c c c c} 
\hline\hline 
Model & Trajectories & Mapping Function & Observation & Single-trial$^{1}$ \\ [0.5ex] 
\hline 
PCA/FA \cite{roweis1999unifying} & Static & Linear & Gaussian & No \\ 
dPCA \cite{kobak2016demixed} & Static & Linear & Gaussian$^{2}$ & No \\ 
jPCA \cite{churchland2012neural} & Linear$^{3}$ & Linear & Gaussian & No \\
LDS \cite{smith2003estimating} & Linear & Linear & Gaussian  & Yes \\
PLDS \cite{macke2012empirical} & Linear & Linear & Poisson  & Yes  \\
PSID \cite{saniModelingBehaviorallyRelevant2020} & Linear & Linear & Gaussian$^{4}$ & Yes \\
PfLDS \cite{gao2016linear} & Linear & Neural Network & Poisson & Yes \\
SLDS \cite{petreskaDynamicalSegmentationSingle2011} & Switching Linear & Linear & Gaussian & Yes \\
RSLDS \cite{linderman2017bayesian, zoltowski2020general} & Recurrent Switching Linear$^{5}$  & Linear & Gaussian & Yes \\
PLRNN-SSM \cite{durstewitz2017state} & Piecewise-linear RNN & Linear & Gaussian & Yes \\
LFADS \cite{pandarinathInferringSingletrialNeural2018} & RNN & Linear & Poisson & Yes \\
GP-RNN \cite{she2020neural} & RNN & GP & Poisson/Gaussian & Yes \\
GPFA \cite{byron2009gaussian} & GP & Linear & Gaussian & Yes\\
GPFADS \cite{rutten2020non} & GP$^{6}$ & Linear & Gaussian & Yes \\ [1ex] 
vLGP \cite{zhao2017variational} & GP & Linear & Poisson$^{7}$ & Yes\\ 
P-GPLVM \cite{wu2017gaussian} & GP & GP & Poisson & Yes \\ 
\hline 
\end{tabular}
\caption{Latent Variable Models: A table (adapted from \cite{she2020neural}) characterizing the latent variable models referenced in this review. 1. The ability to extract dynamics from single-trials is important for capturing important variability that can be obscured with trial-averaging \cite{nawrot1999single}. 2. dPCA models trial-averaged and task averaged neural activity to demix different sources of variability in the data. 3. jPCA is equivalent to a linear dynamical model with specific constraints for modeling rotational dynamics \cite{nemati2014probabilistic}. 4. PSID models both the observed neural activity and behaviour with separate Gaussian observation models. 5. RSLDS can model multiple neural populations and their time-varying interactions using switching linear dynamics that is governed by a discrete latent state. 6. GPFADS introduces a new GP kernel which allows for modeling the dynamical behaviour of temporal-irreversibility. 7. The activity of each neuron in vLGP is linearly dependent on the latent variable and the self-history of the neuron.} 
\label{table:lvm} 
\end{table}

\section*{Outlook}

While both fully observed and latent variable models can be used to overcome the curse of dimensionality in large-scale recordings, the two classes yield distinct yet complementary insights into circuit function. In addition, they each have important limitations that are not easily resolved. Fully observed models can suffer from poor parameter identifiability because data sets have a limited size and potentially because many parameters are actually not well constrained in real neural circuits. Possible future direction are to either simplify the description of the interactions to reduce the number of parameters \cite{rene2020inference} or to augment a model with biologically informed latent dynamics \cite{rule2019neural}. 

Latent variable models assume that circuit dynamics are inherently low-dimensional and place specific constraints on how these dynamics unfold. It is perhaps for this reason that we have, so far, few insights into neural dynamics during complex naturalistic behavior; experimental design still favors simple, repeated stimuli and behaviors where sources of neural variability are easy to de-mix. In more complex experimental regimes, it may no longer be sufficient to generically model the response structure of a neural population and perform ex post facto analyses to relate it back to the variables of interest. Instead, there is increasing evidence that approaches that explicitly model the measured external variables have improved interpretability and can reveal novel insights about neural function \cite{saniModelingBehaviorallyRelevant2020, zhou2020learning}. 

\section*{Acknowledgments}

AO is supported by the Engineering and Physical Sciences Research Council (EP/ S005692/1). CLH is supported by the Thouron Award and by the School of Informatics, University of Edinburgh.

\bibliography{main_final}

\end{document}